\documentclass[journal]{IEEEtran}

\usepackage{color}
\usepackage{float}
\usepackage{graphicx}
\usepackage[position=top, caption=false]{subfig}
\usepackage{mathtools}
\usepackage[utf8]{inputenc}
\usepackage{gensymb}
\usepackage{cite}
\usepackage{hyperref}

\usepackage{mathrsfs}

\begin{document}

\title{Simultaneous Control of the Spatial\\ and Temporal Spectra of Light\\ with Space-Time Varying Metasurfaces}

\author{Nima~Chamanara,
        Yousef~Vahabzadeh,~\IEEEmembership{Student Member,~IEEE,}
        Christophe~Caloz,~\IEEEmembership{Fellow,~IEEE}% <-this % stops a space
\thanks{Polytechnique Montr\'{e}al, Montr\'{e}al, Qu\'{e}bec H3T 1J4, Canada.}% <-this % stops a space
% \thanks{Manuscript received April 10, 2005; revised August 26, 2015.}
}

\markboth{IEEE TRANSACTIONS ON ANTENNAS AND PROPAGATION ,~Vol.~XX, No.~X, August~20XX}%
{Spatio-Temporal Spectrum Control with Space-Time Varying Metasurfaces}

\maketitle

\begin{abstract}
This paper presents space-time varying (STV) metasurfaces for simultaneously controlling the spatial and temporal spectra of electromagnetic waves. These metasurfaces transform incident electromagnetic waves into specified reflected and transmitted waves, with arbitrary temporal and spatial frequencies. They are synthesized in terms of time-domain generalized sheet transition conditions (GSTCs). Moreover, they are characterized using  an analytical method and the unstaggered finite-difference time-domain (FDTD) technique adapted to space-time metasurfaces. STV metasurfaces performing pulse shaping, time reversal and differentiation are demonstrated as examples.
\end{abstract}


\section{Introduction}

Metasurfaces are arrangements of sub-wavelength scattering particles in subwavelengthly-thin sheets, whose geometry and constituents are engineered for specific field transformations~\cite{holloway2012overview, achouri2015general}. They provide unprecedented control over electromagnetic waves and have found many practical applications, such as flat lenses~\cite{yu2014flat}, high-performance holograms~\cite{zheng2015metasurface}, orbital angular momentum generators~\cite{karimi2014generating}, efficient refractive structures~\cite{pfeiffer2014high}, remote field controllers~\cite{achouri2016metasurfaceremote}, light extraction efficiency enhancers~\cite{chen2017simultaneous} and solar sails~\cite{achouri2017solarsail}.

Although metasurface technology has experienced major advances in the past few years, the fundamental option of making them time-variant has been almost unexplored. Given that linear time invariant (LTI) media are limited by bounds imposed by physical laws on static media, the exploration of dynamic metasurfaces, which are free of such limitations, may provide new functionality unobtainable in the realm of static metasurfaces. Such physical limitations involve limits on dispersion engineering imposed by Kramers-Kronig relations, related to causality, or bounds on impedance matching bandwidth of LTI systems imposed by the Bode-Fano criterion~\cite{bode1945network, shlivinski2018paradigm}.

Introducing time variation into metasurfaces both lifts such physical limitations, and provides more degrees of freedom for controlling electromagnetic waves. Moreover space-time variation can naturally break Lorentz reciprocity~\cite{shaltout2015time, hadad2015space, caloz2018nonreciprocity, liu2018huygens}. Such an approach to nonreciprocity has spurred many research advances in the field of magnetless nonreciprocity~\cite{yu2009complete, sounas2013giant, chamanara2017optical, caloz2018nonreciprocity, taravati2017nonreciprocal}, with applications in circulators and isolators, elements for full-duplex communication in microwave and optical systems, circuit protection, matching and stability against back reflections in lasers, radars and radio systems.

This paper leverages space-time variation for general control of the spatio-temporal spectrum of electromagnetic waves. It proposes a systematic approach to design the corresponding susceptibilities for precisely transforming electromagnetic waves in terms of their shape, frequency contents and direction of propagation, over a wide bandwidth. Time reversal and differentiation are presented as examples. With the emergence of tunable materials such as graphene~\cite{neto2009electronic, geim2010rise,  chamanara2013coupler, chamanara2012transparentgraphene, chamanara2016grapheneTEplasmon, chamanara2015fundamentals} or highly tunable transparent conductive oxides~\cite{kinsey2015epsilon, ferrera2017dynamic}, these dynamic metasurfaces may provide features complementary to those of static metasurfaces in the near future.

The organization of the paper is as follows. Section~\ref{sec:principle} presents the principle of operation of STV metasurfaces. Section~\ref{sec:design} derives a method for designing such metasurfaces. Section~\ref{sec:energy-conservation} derives energy relations and verifies energy conservation. Section~\ref{sec:analysis} develops analytical and numerical techniques for characterizing STV metasurfaces. Section~\ref{sec:examples} demonstrates some applications. Finally conclusions are given in Sec.~\ref{sec:conclusions}.

\section{Principle} \label{sec:principle}

\begin{figure}[ht!]
\centering
\includegraphics[page=1]{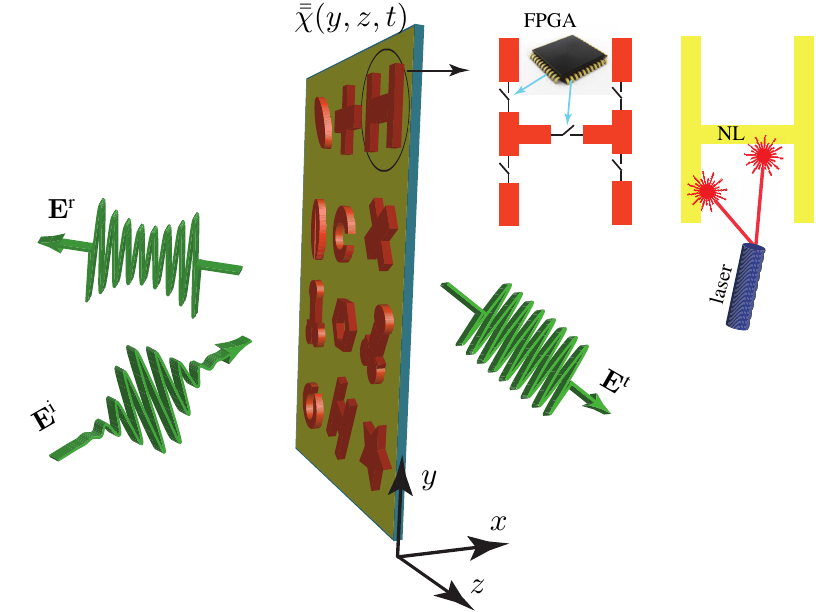}
\caption{Space-time varying (STV) metasurface for simultaneously controlling the temporal and spatial spectra of electromagnetic waves. The metasurface elements are modulated in space and time. Such a dynamic control may be achieved through fast swithing circuits in the microwave regime or through tunable materials such as graphene or transparent conducting oxides or nonlinear media in the optical regime. The metasurface transforms a specified incident wave into arbitrary specified reflected and transmitted waves.}
\label{fig:st-metasurface-concept}
\end{figure}

Figure~\ref{fig:st-metasurface-concept} presents a schematic of the proposed metasurface. It is composed of scattering particles whose parameters are modulated in space and time. Such dynamic control can be achieved in the microwave regime through fast switching circuits, and in the optical regime through nonlinear media excited by strong modulating lasers or through electrically tunable materials such as graphene. The resulting metasurfaces can be described in terms of effective STV susceptibilities.

To understand how STV metasurfaces work, it is instructive to first examine the principles of their time-invariant counterparts. Consider a linear time-invariant (LTI) metasurface illuminated by a \emph{monochromatic wave} with angular frequency $\omega_0$. Since the structure is LTI, different frequencies can be treated independently. The incident wave induces effective electric or magnetic polarization currents on the metasurface according to electromagnetic polarizability of its scattering particles at $\omega_0$. The effective polarization currents then radiate waves corresponding to the scattered fields~\cite{achouri2015general, achouri2018design}. 

The interaction of a LTI metasurface with a \emph{wide band} pulse follows Fourier principles~\cite{chamanara2017efficient, chamanara2016exact}. Each frequency component of the pulse, $\omega_\text{i}$, sees the response of the scattering particles only at the specific frequency $\omega_\text{i}$. Therefore, to achieve a wide-band transformation, the response of the scattering particles must be engineered over the entire bandwidth of the incident pulse. However, the frequency response of LTI metasurfaces is governed by Kramers-Kronig relations~\cite{schwinger1998classical, jackson2012classical, ishimaru2017electromagnetic}. According to these relations, unconventional dispersion is necessarily accompanied by loss. Moreover, the metasurface cannot be perfectly matched to a general load, except at a few discrete frequency points. This limitation corresponds to the Bode-Fano criterion for LTI systems.

STV metasurfaces may potentially break these limits. Although such metasurfaces are still described by linear differential equations, they do not follow similar Fourier principles, since the frequency components get coupled through the STV modulation. The response of STV metasurfaces is best understood by considering their instantaneous interaction with the electromagnetic fields and the modulation, in the time domain.

In the STV metasurfaces presented here, the fields scattered from the metasurface are engineered to follow the temporal evolution of the specifications. Any scattered field can be modeled by equivalent field distributions at $x=0^+$ and $x=0^-$, and the metasurface is controlled in space and time to create such equivalent fields at any instant $t$. Susceptibilities of the metasurface, are dynamically modulated such that, at any instant $t$, the polarization currents follow the temporal profiles dictated by the specified transformations. The required temporal profile of the polarization currents corresponding to the specified transformation are obtained by applying the time-varying form of the generalized sheet transition conditions (GSTCs)~\cite{idemen1987boundary, achouri2015general, achouri2018design}, which relate the discontinuity in the fields to the effective polarization currents on the metasurface at any instant~$t$.

\section{STV Metasurface Theory} \label{sec:design}

Consider an arbitrary STV field transformation, as depicted in Fig.~\ref{fig:st-metasurface-concept}, where an incident pulse is transformed into reflected and transmitted pulses with specified temporal wave forms and scattered into different directions. The corresponding GSTCs read~\cite{idemen1987boundary, achouri2015general, achouri2018design}

\begin{subequations} \label{eq:gstcs-general}
\begin{align}
{\bf{n}} \times \Delta {\bf{H}} &= {\bf{J}} + \frac{{\partial {{\bf{P}}_\text{t}}}}{{\partial t}} - {\bf{n}} \times \nabla {M_\text{n}},\\
{\bf{n}} \times \Delta {\bf{E}} &=  - {\bf{K}} - {\mu_{0}}\frac{{\partial {{\bf{M}}_\text{t}}}}{{\partial t}} - \frac{1}{{{\varepsilon _0}}}{\bf{n}} \times \nabla {P_\text{n}}.
\end{align}
\end{subequations}
These relations connect the equivalent surface electric and magnetic current densities, to the discontinuity of the electromagnetic field across the metasurface. $\mathbf{J}$, $\mathbf{K}$, $\mathbf{P}$ and $\mathbf{M}$ are the electric and magnetic current densities and the electric and magnetic polarization densities, respectively, and $\Delta {\bf{E}} = \bf{E}^+ - \bf{E}^-$, where $+$ and $-$ refer to the fields at $x=0^+$ and $x=0^-$, respectively, $\mathbf{n}$ is the unit vector normal to the metasurface, and the subscripts $\text{n}$ and $\text{t}$ refer to normal and tangential directions with respect to the metasurface components. 

For simplicity~\cite{achouri2018design}, we assume that the metasurface supports only tangential polarization currents. In this case, \eqref{eq:gstcs-general} reduce to

\begin{subequations} \label{eq:gstcs-PM_t_only}
\begin{align}
\mathbf{n}\times\left[\mathbf{H}^{+}(\boldsymbol{\rho}, t)-\mathbf{H}^{-}(\boldsymbol{\rho}, t)\right]&=\frac{\partial}{\partial t}\mathbf{P}_\text{t}(\boldsymbol{\rho}, t),\\
\mathbf{n}\times\left[\mathbf{E}^{+}(\boldsymbol{\rho}, t)-\mathbf{E}^{-}(\boldsymbol{\rho}, t)\right]&=-\mu_{0}\frac{\partial}{\partial t}\mathbf{M}_\text{t}(\boldsymbol{\rho}, t),
\end{align}
\end{subequations}
where $\boldsymbol{\rho} = (0,y,z)$ represents an arbitrary point on the metasurface. Integrating these relations yields the surface polarization densities required for the specified transformation, namely

\begin{subequations} \label{eq:PM-integral}
\begin{align}
\mathbf{P}_\text{t}\left(\boldsymbol{\rho}, t\right)&=\intop_{-\infty}^t\mathbf{n}\times\left[\mathbf{H}^{+}\left(\boldsymbol{\rho}, t^{\prime}\right)-\mathbf{H}^{-}\left(\boldsymbol{\rho}, t^{\prime}\right)\right]dt^{\prime},\\
\mathbf{M}_\text{t}\left(\boldsymbol{\rho}, t\right)&=-\frac{1}{\mu_{0}}\intop_{-\infty}^t\mathbf{n}\times\left[\mathbf{E}^{+}\left(\boldsymbol{\rho}, t^{\prime}\right)-\mathbf{E}^{-}\left(\boldsymbol{\rho}, t^{\prime}\right)\right]dt^{\prime}.
\end{align}
\end{subequations}
Note that the electric and magnetic fields in the right-hand side of \eqref{eq:PM-integral} are assumed to be known from the specified transformation. Therefore $\mathbf{P}_\text{t}\left(\boldsymbol{\rho}, t\right)$ and $\mathbf{M}_\text{t}\left(\boldsymbol{\rho}, t\right)$ can be directly computed by performing the integrals in \eqref{eq:PM-integral}. 

The sought after susceptibilities connect the polarization densities in \eqref{eq:PM-integral} to the average electric and magnetic fields on the metasurface through constitutive relations, which read in the most general (bianisotropic) case

\begin{subequations} \label{eq:constitutive-bianisotropic}
\begin{align}
{\mathbf{P}(\boldsymbol{\rho},t)} &= {\varepsilon _0}{{\bar {\bar \chi}}_\text{ee}(\boldsymbol{\rho},t)} \cdot {\bf{E}_\text{av}} + \sqrt {{\mu _0}{\varepsilon _0}} {{\bar {\bar \chi} }_\text{em}(\boldsymbol{\rho},t)} \cdot {\bf{H}_\text{av}},\\
{\mathbf{M}(\boldsymbol{\rho},t)} &= {{\bar {\bar \chi} }_\text{mm}(\boldsymbol{\rho},t)} \cdot {\bf{H}_\text{av}} + \sqrt {{\varepsilon _0}/{\mu _0}} {{\bar {\bar \chi} }_\text{me}(\boldsymbol{\rho},t)} \cdot {\bf{E}_\text{av}},
\end{align}
\end{subequations}
where the subscript ``av" denote the average fields, given by

\begin{subequations}
\begin{align}
\mathbf{{\mathbf{E}_\text{av}}}\left(\boldsymbol{\rho}, t\right) &= \frac{1}{2}\left[\mathbf{E}^+\left(\boldsymbol{\rho}, t\right) + \mathbf{E}^-\left(\boldsymbol{\rho}, t\right)\right] \\\nonumber
&=\frac{1}{2}\left[\mathbf{E}^\text{i}\left(\boldsymbol{\rho}, t\right)+\mathbf{E}^\text{r}\left(\boldsymbol{\rho}, t\right)+\mathbf{E}^\text{t}\left(\boldsymbol{\rho}, t\right)\right],\\
\mathbf{\mathbf{H}_\text{av}}\left(\boldsymbol{\rho}, t\right) &= \frac{1}{2}\left[\mathbf{H}^+\left(\boldsymbol{\rho}, t\right) + \mathbf{H}^-\left(\boldsymbol{\rho}, t\right)\right] \\\nonumber
&= \frac{1}{2}\left[\mathbf{H}^\text{i}\left(\boldsymbol{\rho}, t\right)+\mathbf{H}^\text{r}\left(\boldsymbol{\rho}, t\right)+\mathbf{H}^\text{t}\left(\boldsymbol{\rho}, t\right)\right].
\end{align}
\end{subequations}

For the sake of simplicity, we restrict ourselves here to diagonal monoanisotropic susceptibilities, i.e.,

\begin{subequations} \label{eq:X-diagonal}
\begin{align}
\bar{\bar{\boldsymbol{\chi}}}_\text{ee} &= \chi_\text{ee}^{zz} \hat{\mathbf{z}}\hat{\mathbf{z}} +  \chi_\text{ee}^{yy} \hat{\mathbf{y}}\hat{\mathbf{y}}, \\
\bar{\bar{\boldsymbol{\chi}}}_\text{mm} &= \chi_\text{mm}^{zz} \hat{\mathbf{z}}\hat{\mathbf{z}} +  \chi_\text{mm}^{yy} \hat{\mathbf{y}}\hat{\mathbf{y}}, \\
\bar{\bar{\boldsymbol{\chi}}}_\text{em} &= \bar{\bar{\boldsymbol{\chi}}}_\text{me} = \mathbf{0},
\end{align} 
\end{subequations}
which, with the above tangential polarization assumption, reduce \eqref{eq:constitutive-bianisotropic} to 

\begin{subequations}
\begin{align}
\mathbf{P}_\text{t}\left(\boldsymbol{\rho}, t\right)&=\varepsilon_{0}\bar{\bar{\boldsymbol{\chi}}}_\text{ee}\left(\boldsymbol{\rho}, t\right)\cdot{\mathbf{E}_\text{av}}\left(\boldsymbol{\rho}, t\right), \\
\mathbf{M}_\text{t}\left(\boldsymbol{\rho}, t\right)&=\bar{\bar{\boldsymbol{\chi}}}_\text{mm}\left(\boldsymbol{\rho}, t\right)\cdot {\mathbf{H}_\text{av}}\left(\boldsymbol{\rho}, t\right).
\end{align} \label{eq:constitut-mono-iso}
\end{subequations}

\noindent
The discussion of more general constitutive relations is deferred to the end of this section. Inserting \eqref{eq:constitut-mono-iso} into \eqref{eq:PM-integral} provides the required STV susceptibilities for a given pulse transformation

\begin{subequations}
\begin{align}
&\bar{\bar{\boldsymbol{\chi}}}_\text{ee}\left(\boldsymbol{\rho}, t\right)= \\\nonumber
&\frac{1}{\varepsilon_{0}} \intop_{-\infty}^t\mathbf{n}\times\left[\mathbf{H}^{+}\left(\boldsymbol{\rho}, t^{\prime}\right)-\mathbf{H^{-}\left(\boldsymbol{\rho}, t^{\prime}\right)}\right]dt^{\prime} ./  \mathbf{{\mathbf{E}_\text{av}}}\left(\boldsymbol{\rho}, t\right), \\
&\bar{\bar{\boldsymbol{\chi}}}_\text{mm}\left(\boldsymbol{\rho}, t\right)= \\\nonumber
&-\frac{1}{\mu_{0}} \intop_{-\infty}^t\mathbf{n}\times\left[\mathbf{E}^{+}\left(\boldsymbol{\rho}, t^{\prime}\right)-\mathbf{E^{-}\left(\boldsymbol{\rho}, t^{\prime}\right)}\right]dt^{\prime}  ./ {\mathbf{H}_\text{av}}\left(\boldsymbol{\rho}, t\right),
\end{align} \label{eq:X-mono-final}
\end{subequations}

\noindent
where $./$ is the \emph{array division} operator, whose resulting $z$ component is the ratio of the $z$ component of the numerator to the $z$ component of the denominator, and $y$ component is the ratio of the $y$ component of the numerator to the $y$ component of the denominator, while the $\mathbf{z}\mathbf{y}$ and $\mathbf{y}\mathbf{z}$ components are zero. 

Equations~\eqref{eq:X-mono-final} can be easily generalized to the most general bianisotropic case by using \eqref{eq:constitutive-bianisotropic} instead of \eqref{eq:constitut-mono-iso}. However, Eq.~\eqref{eq:constitutive-bianisotropic} might lead to an under-determined system, depending on the number of non-zero components in the susceptibility tensors. In that case, one can either introduce extra field transformations to arrive at a fully determined system, or use the extra degrees of freedom to optimize the susceptibilities for a particular goal. For more details the reader is referred to~\cite{achouri2015general, achouri2018design}.

\section{Energy Relations and Conservation}  \label{sec:energy-conservation}

\begin{figure}[ht!]
	\centering
	\includegraphics[page=2]{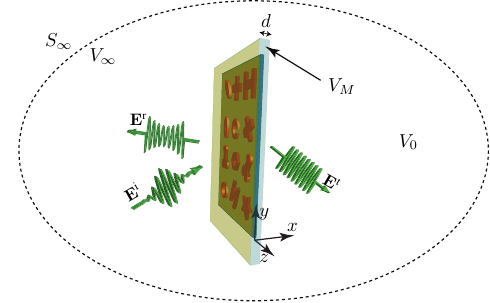}
	\caption{Definition of volumes required to calculate power balance in a STV metasurface transformation. $S_\infty$ represents a surface enclosing an infinitely large fictitious volume, $V_\infty$, encompassing the metasurface and the scattered fields. $V_\text{M}$ represents an infinitesimally thin volume containing the metasurface and $V_0 = V_\infty - V_\text{M}$.}
	\label{fig:st-metasurface-power-balance}
\end{figure}

This section derives energy relations for the problem of scattering from an STV metasurface which is represented in Fig.~\ref{fig:st-metasurface-power-balance} and verifies energy conservation. The metasurface is enclosed in the volume $V_\infty$ delimited by the surface $S_\infty$, which extends to infinity. The volume $V_\text{M}$ represents a thin parallelepiped volume with an infinitesimal thickness $d$ tightly enclosing the metasurface, and $V_0 = V_\infty - V_\text{M}$. 

We first calculate the divergence of the Poynting vector as

\begin{align} \label{eq:poynting-derivation}
-\nabla\cdot\boldsymbol{S}=&-\nabla\cdot\left(\boldsymbol{E}\times\boldsymbol{H}\right)\\ \nonumber
=&-\boldsymbol{H}\cdot\nabla\times\boldsymbol{E}+\boldsymbol{E}\cdot\nabla\times\boldsymbol{H}\\ \nonumber
=&\boldsymbol{H}\cdot\frac{\partial}{\partial t}\boldsymbol{B}+\boldsymbol{E}\cdot\left(\boldsymbol{J}+\frac{\partial}{\partial t}\boldsymbol{D}\right)\\ \nonumber
=&\boldsymbol{E}\cdot\frac{\partial}{\partial t}\boldsymbol{D} + \boldsymbol{H}\cdot\frac{\partial}{\partial t}\boldsymbol{B} + \boldsymbol{E}\cdot\boldsymbol{J},
\end{align}

\noindent
where $\mathbf{J}$ represents enforced and conduction currents. Note that \eqref{eq:poynting-derivation} is general and does not make any assumption on material parameters. 

The terms involving $\mathbf{D}$ and $\mathbf{B}$ in \eqref{eq:poynting-derivation} are then separately calculated in regions $V_0$ and $V_\text{M}$, leading to

\begin{align} \label{eq:em-energy}
\boldsymbol{E}\cdot\frac{\partial}{\partial t}\boldsymbol{D}+\boldsymbol{H}\cdot\frac{\partial}{\partial t}\boldsymbol{B} = \frac{\partial}{\partial t} \mathscr{U} - \mathscr{P}_\text{M},
\end{align}

\noindent
where $\mathscr{U}$ is the electromagnetic energy density in $V_0$, which reads

\begin{align}
\mathscr{U} = \frac{1}{2}\epsilon_{0} \boldsymbol{E}\cdot\boldsymbol{E}+\frac{1}{2}\mu_{0}\boldsymbol{H}\cdot\boldsymbol{H},
\end{align}

\noindent
and $\mathscr{P}_\text{M}$ is the power delivered by the metasurface which is expressed as

\begin{align} \label{eq:P_M_volumetric}
\mathscr{P}_{M}=&-\epsilon_{0}\boldsymbol{E}_\text{av}\cdot\frac{\partial}{\partial t}\left\{ \left[1+\bar{\bar{\chi}}_\text{ee}\delta\left(x\right)\right]\cdot\boldsymbol{E}_\text{av}\right\} \\ \nonumber
&-\mu_{0}\boldsymbol{H}_\text{av}\cdot\frac{\partial}{\partial t}\left\{ \left[1+\bar{\bar{\chi}}_\text{mm}\delta\left(x\right)\right]\cdot\boldsymbol{H}_\text{av}\right\}.
\end{align}

In \eqref{eq:poynting-derivation} the term involving $\mathbf{J}$ can be expanded as

\begin{align}
\boldsymbol{E}\cdot\boldsymbol{J}=\boldsymbol{E}\cdot\boldsymbol{J}_\text{s}+\sigma\boldsymbol{E}\cdot\boldsymbol{E},
\end{align}

\noindent
where $\boldsymbol{J}_\text{s}$ is the enforced current source and $\sigma$ is the conductivity inside $V_\infty$.

The energy conservation relation finally reads

\begin{subequations}
\begin{align} \label{eq:energy-conservation-local}
\frac{\partial}{\partial t}\mathscr{U} &= \mathscr{P}_\text{M} - \nabla\cdot\boldsymbol{S} + \mathscr{P}_\text{J} - \mathscr{P}_\text{L}, \\
\mathscr{P}_\text{J} &=-\boldsymbol{E}\cdot\boldsymbol{J}_{s} \\
\mathscr{P}_\text{L} &=\sigma\boldsymbol{E}\cdot\boldsymbol{E}
\end{align}
\end{subequations}

\noindent
where $\mathscr{P}_\text{J}$ is the power delivered by source currents and $\mathscr{P}_\text{L}$ is the power lost in heat. This equation states that the time rate of change in the electromagnetic energy density, $\mathscr{U}$, equals the sum of power delivered by the metasurface, $ \mathscr{P}_\text{M}$,  and the enforced currents, $\mathscr{P}_\text{J}$, minus the power that is dissipated in heat, $\mathscr{P}_\text{L}$, and the power that propagates out of $S_\infty$. In what follows we assume that $\boldsymbol{J}_\text{s} = \mathscr{P}_\text{J} = 0$, and that the metasurface is composed of lossless media, i.e., $\mathscr{P}_\text{L} = 0$. 

Equation \eqref{eq:energy-conservation-local} integrates as

\begin{align} \label{eq:energy-conservation-integral}
\frac{\partial}{\partial t}\int_{V_{0}} \mathscr{U} dv=\int_{V_{M}} \mathscr{P}_{M}dv-\int_{S_{\infty}}\boldsymbol{S}\cdot d\boldsymbol{s}.
\end{align}

We assume that the incident wave is a pulse that is finite in both space and time and the surface $S_\infty$ is at infinity, so that the scattered waves never cross $S_\infty$. We also let $d \rightarrow 0$ so that only the terms involving the delta functions in \eqref{eq:P_M_volumetric} contribute to the volume integral in \eqref{eq:energy-conservation-integral}. The volume integral of $\mathscr{P}_\text{M}$ reduces then to a surface integral over the metasurface, and \eqref{eq:energy-conservation-integral} becomes

\begin{align} \label{eq:energy-conservation-integral-P_M-only}
\frac{\partial}{\partial t}\int_{V_{0}} \mathscr{U}dv=\int_{S_{M}} \mathscr{P}_{M}ds.
\end{align}

\noindent
This relation states that the change in the electromagnetic energy density is balanced by the power delivered by (for $\mathscr{P}_\text{M}>0$) or absorbed by (for $\mathscr{P}_\text{M}<0$) the metasurface, where $\mathscr{P}_\text{M}$ reads

\begin{align} \label{eq:P_M_on-metasurface}
\mathscr{P}_\text{M}=-\epsilon_{0}\boldsymbol{E}_\text{av}\cdot\frac{\partial}{\partial t}\left(\bar{\bar{\chi}}_\text{ee}\cdot\boldsymbol{E}_\text{av}\right)-\mu_{0}\boldsymbol{H}_\text{av}\cdot\frac{\partial}{\partial t}\left(\bar{\bar{\chi}}_\text{mm}\cdot\boldsymbol{H}_\text{av}\right).
\end{align}

\noindent
The expressions involving time derivatives can be further simplified using \eqref{eq:constitutive-bianisotropic} and \eqref{eq:gstcs-PM_t_only}, reducing \eqref{eq:P_M_on-metasurface} to

\begin{align}
\mathscr{P}_\text{M}=&-\boldsymbol{E}_\text{av}\cdot\left[\hat{\boldsymbol{n}}\times\left(\boldsymbol{H}^{+}-\boldsymbol{H}^{-}\right)\right] \\ \nonumber
&+ \boldsymbol{H}_\text{av} \cdot\left[\hat{\boldsymbol{n}}\times\left(\boldsymbol{E}^{+}-\boldsymbol{E}^{-}\right)\right],
\end{align}

\noindent
which simply involves the average electromagnetic field on the metasurface and the field on either side of it. This relation states that the amount of power required for an STV transformation with finite electromagnetic fields is finite.

\section{Analysis} \label{sec:analysis}
This section presents analytical and numerical techniques for characterizing time varying (TV) and STV metasurfaces. The goal is to find the fields scattered by the metasurface, for given susceptibilities and incident field, in order to validate the synthesis performed in Sec.~\ref{sec:design}. When the incident, reflected and transmitted fields are plane waves that are normal to the metasurface, the susceptibilities are \emph{purely time-varying}, i.e., include no spatial variation. In this case, the problem can be analyzed using analytical techniques. For the more general case of STV metasurfaces, these analytical techniques becomes intractable, and we therefore resort to numerical simulation using the finite-difference time-domain (FDTD) technique~\cite{taflove2005computational}.

\subsection{1D Time-Varying Metasurfaces}

Consider a normally incident $z$-polarized plane wave \mbox{$\mathbf{E}^\text{i}(\mathbf{r}, t) = E^\text{i}(t - \frac{x}{c})\hat{\mathbf{z}}$} impinging on a metasurface with purely time-varying isotropic electric and magnetic susceptibilities. The GSTCs \eqref{eq:gstcs-PM_t_only} reduce then to

\begin{subequations}
\begin{align}
\frac{\partial}{\partial t}P_\text{t}\left(t\right)=-\frac{1}{\eta_{0}}\left[-E^\text{i}\left(t\right)+E^\text{r}\left(t\right)+E^\text{t}\left(t\right)\right], \\
-\mu_{0}\frac{\partial}{\partial t}M_\text{t}\left(t\right)=\left[-E^\text{i}\left(t\right)-E^\text{r}\left(t\right)+E^\text{t}\left(t\right)\right],
\end{align} \label{eq:time-only-1d-gstc-mono-iso}
\end{subequations}
and the constitutive relation \eqref{eq:constitut-mono-iso} reduce to

\begin{subequations}
\begin{align}
\frac{1}{2}\varepsilon_{0}\mathbf{\chi}_\text{ee}\left(t\right)\left[E^\text{i}\left(t\right)+E^\text{r}\left(t\right)+E^\text{t}\left(t\right)\right]=P\left(t\right), \\
\frac{1}{2\eta_{0}}\mathbf{\chi}_\text{mm}\left(t\right)\left[E^\text{i}\left(t\right)-E^\text{r}\left(t\right)+E^\text{t}\left(t\right)\right]=M\left(t\right).
\end{align} \label{eq:time-only-1d-gstc-mono-iso-xpolarized}
\end{subequations}
Equation~\eqref{eq:time-only-1d-gstc-mono-iso-xpolarized} is then solved for $E^\text{r}$ and $E^\text{t}$

\begin{subequations}
\begin{align}
E^\text{r}\left(t\right) &= \frac{P_\text{t}\left(t\right)}{\varepsilon_{0}\mathbf{\chi}_\text{ee}\left(t\right)} - \eta_{0} \frac{M_\text{t}\left(t\right)}{\mathbf{\chi}_\text{mm}\left(t\right)}, \\
E^\text{t}\left(t\right) &= \frac{P_\text{t}\left(t\right)}{\varepsilon_{0}\mathbf{\chi}_\text{ee}\left(t\right)} + \eta_{0} \frac{M_\text{t}\left(t\right)}{\mathbf{\chi}_\text{mm}\left(t\right)} - E^\text{i}\left(t\right).
\end{align} \label{eq:time-only-1d-gstc-mono-iso-ErEt}
\end{subequations}
Substituting these expressions into \eqref{eq:time-only-1d-gstc-mono-iso}, results in

\begin{subequations}
\begin{align}
-\eta_{0}\frac{\partial}{\partial t}P_\text{t}\left(t\right)=\frac{2P_\text{t}\left(t\right)}{\varepsilon_{0}\mathbf{\chi}_\text{ee}\left(t\right)}-2E^\text{i}\left(t\right), \\
-\mu_{0}\frac{\partial}{\partial t}M_\text{t}\left(t\right)=\frac{2\eta_{0} M_\text{t}\left(t\right)}{\mathbf{\chi}_\text{mm}\left(t\right)}-2E^\text{i}\left(t\right).
\end{align} \label{eq:time-only-1d-gstc-mono-iso-PM}
\end{subequations}
Note that Eqs.~\ref{eq:time-only-1d-gstc-mono-iso-PM} represent \emph{decoupled} electric and magnetic equations. This is not surprising since the magneto-electric susceptibilities, that produce electric--magnetic coupling effects, were assumed to be zero. Equations \eqref{eq:time-only-1d-gstc-mono-iso-PM} are linear first-order differential equations, which are integrable. Their analytical solutions are provided in Appendix~\ref{appendix:analytic-1d}. The polarization densities $P_\text{t}$ and $M_\text{t}$ are then substituted into \eqref{eq:time-only-1d-gstc-mono-iso-ErEt} to provide the reflected and transmitted fields, $E^\text{r}(x=0, t)$ and $E^\text{t}(x=0, t)$, on the metasurface. The reflected and transmitted fields at any other point $x$ are then obtained as $E^\text{r}(x, t) = E^\text{r}(0, t + x/c)$ and $E^\text{t}(x, t) = E^\text{t}(0, t - x/c)$, respectively.

\subsection{2D STV metasurfaces} \label{Sec:unstaggered-fdtd}

Scattering from general STV metasurfaces can be numerically simulated using FDTD. However, such metasurfaces are not compatible with the conventional staggered (Yee) grid. In the staggered grid, the electric and magnetic currents are placed at electric and magnetic nodes, respectively. As metasurfaces generally produce both electric and magnetic polarization currents, placing a metasurface in this grid would introduce a distance of half a unit cell between the electric and magnetic metasurface currents. Moreover, the electric and magnetic nodes are also staggered in time, which introduces undesired delays between the electric and magnetic metasurface currents. These inaccuracies would produce unphysical errors in the scattered fields. For example a matched metasurface with equal isotropic electric and magnetic susceptibilities would appear as mismatched to a normal incident field by an erroneous amount that is proportional to the cell size, due to these staggered grid effects. Eliminating such errors would require dramatically reducing the mesh size.

Introducing \emph{virtual nodes} between the electric and magnetic nodes, and placing the metasurface at such virtual nodes, reduces the error~\cite{vahabzadeh2018generalized, vahabzadeh2018computational}. However, virtual nodes introduce extra complexity in the FDTD formulation, as they require modifying the conventional update equations according to GSTCs. 

\begin{figure}[ht!]
\centering
\subfloat[~~~~~~~~~~~~~~~~~~~~~~~~~~~~~~~~~~~~~~~~~~~~~~~~~~~~~~~~~~~~~~~~~~~~~~]{ \label{fig:fdtd-unstaggered-grid-st-1d}
\includegraphics[page=3]{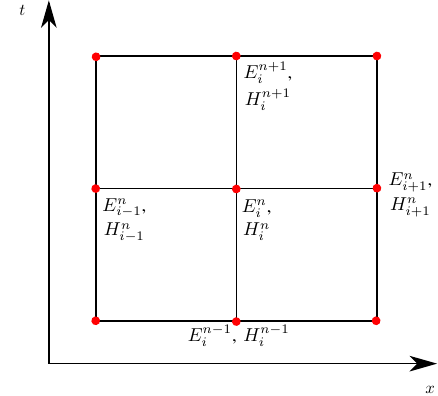} 
}
\vfill
\subfloat[~~~~~~~~~~~~~~~~~~~~~~~~~~~~~~~~~~~~~~~~~~~~~~~~~~~~~~~~~~~~~~~~~~~~~~]{ \label{fig:fdtd-staggered-grid-st-1d}
\includegraphics[page=4]{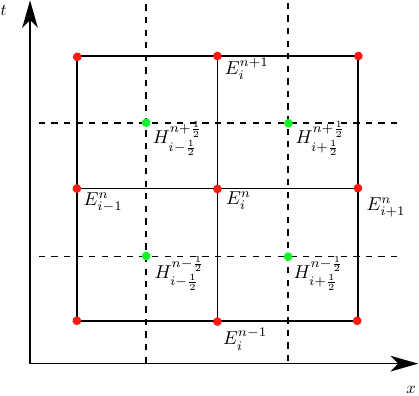} 
}
\caption{ Comparison of the unstaggered and staggered (Yee) FDTD grids in the space-time diagram. (a) Unstaggered grid. The electric and magnetic fields are located at the same grid points and updated at the same times. (b) Staggered grid. The electric and magnetic fields are staggered in space and in time. }
\label{fig:fdtd-grid-st-1d}
\end{figure}

To avoid these problems, we put here the electric and magnetic metasurface currents at the \emph{same} nodes and therefore use the \emph{unstaggered} FDTD technique~\cite{gilles2000comparison, janaswamy1997unstaggered} in the simulation of STV metasurfaces. In this scheme, the electric and magnetic fields are placed at the same nodes and are updated at the same time. The unstaggered FDTD scheme is represented in Fig.~\ref{fig:fdtd-unstaggered-grid-st-1d} in a space-time (Minkowski) diagram, where the subscripts and superscripts represent spatial and temporal indices, respectively\footnote{ Note that the topic computation analysis of metasurfaces has received considerable attention recently, particularly on FDTD modelling of STV metasurfaces~\cite{stewart2018finite, vahabzadeh2018computational, vahabzadeh2018generalized}. For a comprehensive review of this topic, the reader is referred to \cite{vahabzadeh2018computational}.}.

Figure~\ref{fig:fdtd-grid-st-1d} compares the unstaggered and staggered FDTD space-time grids for a 1D problem. In the unstaggered grid, shown in Fig.~\ref{fig:fdtd-unstaggered-grid-st-1d}, the electric and magnetic fields are placed at the same nodes in space and time and are updated at the same time instants. In contrast, in the staggered (Yee) grid, shown in Fig.~\ref{fig:fdtd-staggered-grid-st-1d}, the electric and magnetic quantities are staggered both in space and in time. 

\begin{figure}[ht!]
\centering
\subfloat[~~~~~~~~~~~~~~~~~~~~~~~~~~~~~~~~~~~~~~~~~]{ \label{fig:fdtd-unstaggered-grid-xy-2d}
\includegraphics[page=5]{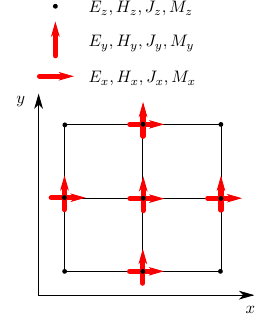} 
}
\hfill
\subfloat[~~~~~~~~~~~~~~~~~~~~~~~~~~~~~~~~~~~~~~~~~]{ \label{fig:fdtd-staggered-grid-xy-2d}
\includegraphics[page=6]{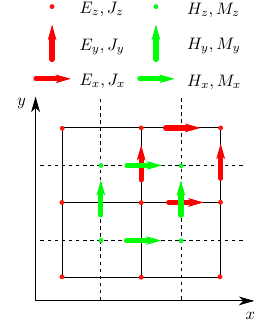} 
}
\caption{ Comparison of the location of the different field and current components in the unstaggered and staggered grids in the 2D $xy$-plane. (a) Unstaggered grid. All the field/current components are colocated at the nodes. (b) Staggered (Yee) grid. The $x$ and $y$ components are located on the horizontal and vertical sides, respectively, and the $z$ components are located at the nodes.}
\label{fig:fdtd-grid-xy-2d}
\end{figure}

Figure~\ref{fig:fdtd-grid-xy-2d} compares the location of the field components in the unstaggered and staggered grids. Note that these graphs represent the $xy$-plane for a two-dimensional problem. In the unstaggered grid, shown in Fig.~\ref{fig:fdtd-unstaggered-grid-xy-2d}, the $x, y, z$ components of all the fields are placed at the nodes. In contrast, in the two-dimensional staggered Yee grid, shown in Fig.~\ref{fig:fdtd-staggered-grid-xy-2d}, the $x$ components are placed at the horizontal sides, the $y$ components at the vertical sides and the $z$ components at the nodes.

In the unstaggered grid the spatial and temporal derivatives are expressed in terms of central difference schemes, 

\begin{subequations}
\begin{align}
\frac{\partial}{\partial t}f\left(x, y, t\right) \approx&  \left[f\left(x, y, t+\Delta t\right) - f\left(x, y, t-\Delta t\right)\right]/ \left(2\Delta t\right), \\
\frac{\partial}{\partial x}f\left(x, y, t\right) \approx&  \left[f\left(x+\Delta x, y, t\right) - f\left(x-\Delta x, y, t\right)\right]/\left(2\Delta x\right), \\
\frac{\partial}{\partial y}f\left(x, y, t\right) \approx&  \left[f\left(x, y+\Delta y, t\right) - f\left(x, y-\Delta y, t\right)\right]/\left(2\Delta y\right).
\end{align} \label{eq:fdtd-dxdt-central-difference}
\end{subequations}
The stability conditions of this scheme are identical to those of the staggered (Yee) grid, and its error is similarly proportional to the square of the grid resolution~\cite{liu1996fourier}. However, it is clear from Figs.~\ref{fig:fdtd-grid-st-1d} and \ref{fig:fdtd-grid-xy-2d} that the resolution of the staggered grid is twice that of the unstaggered one. Therefore, for the same error level, the mesh size of the unstaggered grid should be half that of the staggered grid. It should be noted though, that it is possible to reach the same level of accuracy as the staggered grid, with the same mesh size, using more advanced finite differencing schemes compared to the simple central differencing in \eqref{eq:fdtd-dxdt-central-difference}~\cite{liu1996fourier}. More details on the discretization of Maxwell equations based on \eqref{eq:fdtd-dxdt-central-difference} are provided in Appendix~\ref{appendix:fdtd-unstaggered-maxwell}.

\section{Examples}   \label{sec:examples}

\begin{figure}[ht!] 
\centering
\subfloat[~~~~~~~~~~~~~~~~~~~~~~~~~~~~~~~~~~~~~~~~~]{ \label{fig:metasurf_TR_T}
\includegraphics[page=7]{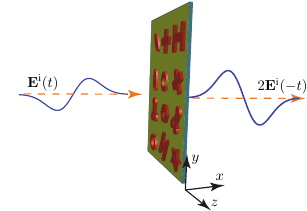}
}
\vfill
\subfloat[~~~~~~~~~~~~~~~~~~~~~~~~~~~~~~~~~~~~~~~~~]{ \label{fig:metasurf_TR_ST}
\includegraphics[page=8]{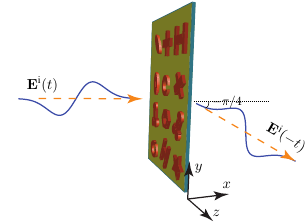}
}
\vfill
\subfloat[~~~~~~~~~~~~~~~~~~~~~~~~~~~~~~~~~~~~~~~~~]{ \label{fig:metasurf_TR_diff_ST}
\includegraphics[page=9]{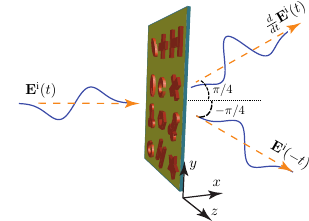}
}
\caption{STV metasurface examples. (a) 1D time-reversing amplifying STV metasurface. (b) Time-reversal refractive space-time metasurface. (c) Multifunction time-reversing differentiating STV metasurface. }
\label{fig:metasurf_ST_examples}
\end{figure}

This section presents examples of STV metasurface of increasing complexity. The examples are broadly presented in Fig.~\ref{fig:metasurf_ST_examples}. Figure~\ref{fig:metasurf_TR_T} depicts a TV metasurface that time-reverses and amplifies a normally incident (asymmetric) pulse. Figure~\ref{fig:metasurf_TR_ST} depicts a STV metasurface that time-reverses an incident pulse and refracts it obliquely, and hence transforms both its temporal and spatial spectra. Finally, Fig.~\ref{fig:metasurf_TR_diff_ST} depicts a STV multifunction metasurface that time-reverses and differentiates an incident pulse and refracts the results at different oblique angles. Details on each transformation are provided below. In all cases, the metasurface is assumed to be monoisotropic ($\bar{\bar{\boldsymbol{\chi}}}_\text{ee} = \chi_\text{ee}$, $\bar{\bar{\boldsymbol{\chi}}}_\text{mm} = \chi_\text{mm}$, $\bar{\bar{\boldsymbol{\chi}}}_\text{em} = \bar{\bar{\boldsymbol{\chi}}}_\text{me} = 0$).

\subsection{1D Time-Reversal Metasurface}

\begin{figure}[ht!]
	\centering
	\subfloat[~~~~~~~~~~~~~~~~~~~~~~~~~~~~~~~~~~~~~~~~~]{ \label{fig:E-TR-irt-1d}
			\includegraphics[page=10]{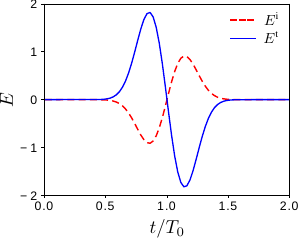}
	}
	\vfill
	\subfloat[~~~~~~~~~~~~~~~~~~~~~~~~~~~~~~~~~~~~~~~~~]{ \label{fig:X-TR-1d}
			\includegraphics[page=11]{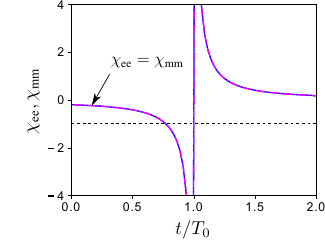}
	}
	\vfill
	\subfloat[~~~~~~~~~~~~~~~~~~~~~~~~~~~~~~~~~~~~~~~~~]{ \label{fig:P-TR-1d}
			\includegraphics[page=12]{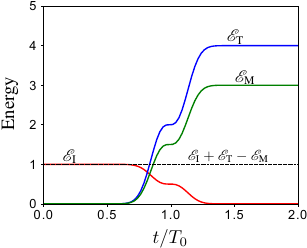}
	}
	\caption{1D time-reversing and amplifying metasurface. (a)~Specified incident and transmitted fields on the metasurface (z=0). (b)~Corresponding TV electric and magnetic susceptibilities based on \eqref{eq:X-mono-final}. (c) Normalized incident energy, $\mathscr{E}_\text{I}$, transmitted energy, $\mathscr{E}_\text{I}$, and energy delivered by the metasurface, $\mathscr{E}_\text{M}$. $T_0$ is the width of the pulse.}
	\label{fig:TR-1d-Es-and-Xs}
\end{figure}

Consider an asymmetric plane-wave pulse impinging normally on the TV metasurface in Fig.~\ref{fig:metasurf_TR_T}. The metasurface is designed to time-reverse and amplify this pulse and then transmit it normally without any reflection. The incident and transmitted pulses are represented in Fig.~\ref{fig:E-TR-irt-1d} in red and blue, respectively, on the metasurface i.e., at $x=0$. Here the fields and the metasurface susceptibilities have no dependence on the $z$ and $y$ coordinates (since all the fields are normal plane waves). The asymmetric incident pulse is given by

\begin{subequations}
\begin{align} \label{eq:Ei-waveform}
	E^\text{i}(\mathbf{r}, t) &= E_0 f\left(t - \frac{x}{c}\right) \hat{\mathbf{z}} \\
	f(\zeta) &= \exp\left[-\left(\zeta - \zeta_0\right)^2/\tau^2\right]
	- \exp\left[-\left(\zeta + \zeta_0\right)^2/\tau^2\right]
\end{align}
\end{subequations}

\noindent
where $\zeta_0 = T_0/4$ and $\tau = 3T_0$, corresponding to a pulse with temporal width $T_0$.

The time-varying susceptibilities, computed by \eqref{eq:X-mono-final}, are plotted in Fig.~\ref{fig:TR-1d-Es-and-Xs}. Note that the electric and magnetic susceptibilities are equal. This is expected, since the metasurface is designed to be matched at all times, corresponding to the impedance $\eta(t)=\eta_0\sqrt{\frac{1 + \chi_\text{mm}(t)}{1 + \chi_\text{ee}(t)}}=\eta_0$. Note that this is the conventional Huygens matching condition expressed in terms of impedances instead of polarizabilities. The singularity corresponds to the center point of the pulse. Close to this point, the average field on the metasurface approaches zero, and therefore the metasurface has to provide infinitely large susceptibilities to produce the required polarization densities out of a vanishingly small field.

The amplification is provided by the pump. Note that in time varying systems, electromagnetic energy is not conserved. The fields can be amplified or attenuated without employing any gain or loss material. The extra energy is provided to or extracted from the system through the pump mechanism. Figure~\ref{fig:P-TR-1d} verifies energy balance. The red and blue curves represent the total instantaneous energy on the left and right side of the metasurface respectively, and the green curve represents the energy delivered by the metasurface. All values are normalized to the total incident energy. The difference between the transmitted and incident energy is exactly compensated by the energy delivered by the metasurface. Note that although the susceptibilities diverge at the center time, the energy delivered by the metasurface remains finite. This is because the average electromagnetic field converges to zero at the singularity, according to \eqref{eq:P_M_on-metasurface} the power delivered by the metasurface converges to a finite value.

Note that the permittivity and permeability corresponding to the susceptibilities in Fig.~\ref{fig:X-TR-1d} become negative for a short time before the resonance. A \emph{non-dispersive} negative permittivity or permeability would correspond to negative electric or magnetic energies and is therefore nonphysical~\cite{schwinger1998classical, jackson2012classical,  ishimaru2017electromagnetic}. However, these susceptibility parameters have been engineered only for the entire bandwidth of the incident and transformed pulses, and can take any values outside that bandwidth, so that they do not violate energy constraints.

\subsection{2D Time-Reversal Refraction STV Metasurface}

Consider an asymmetric Gaussian pulse normally incident on the time-reversal refractive STV metasurface represented in Fig.~\ref{fig:metasurf_TR_ST}. The metasurface time reverses the pulse and perfectly (without spurious diffraction orders) refracts it at $-45$ degrees. The incident pulse is expressed by the spatio-temporal dependence 

\begin{subequations}
\begin{align} \label{eq:Ei-waveform-tr}
	E(x_0, y, t) = E_0 g\left( y \right)f\left( t \right) \hat{\mathbf{z}} \\
	g\left( y \right) = \exp\left(-y^2/W^2\right)
\end{align}
\end{subequations}

\noindent
at the launching plane $x_0 = -20\lambda_0$, where the function $f(.)$ is given in \eqref{eq:Ei-waveform}. The Gaussian-pulse waist is $W = 5\lambda_0$, where $\lambda_0 = cT_0$ and $\omega_0 = 2\pi/T_0$. Away from the launching plane $x = x_0$, at any other point in the $xy$ plane, the electromagnetic field vectors can be obtained using the plane-wave (spectral) expansion method (see Section 6.5 in~\cite{ishimaru2017electromagnetic}) or numerical techniques such as FDTD. The specified transmission pulse can be obtained from \eqref{eq:Ei-waveform-tr} by first applying time-reversal to $f(.)$, then using plane-wave expansion or FDTD to obtain the fields in the entire $xy$ plane, and finally performing a spatial rotation.

\begin{figure}[ht!]
\centering
\subfloat[~~~~~~~~~~~~~~~~~~~~~~~~~~~~~~~~~~~~~~~~~]{ \label{fig:X_ee-tr-2d}
\includegraphics[page=13]{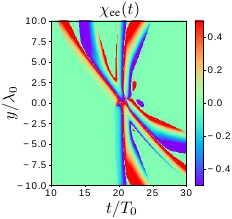}
}
\hfill
\subfloat[~~~~~~~~~~~~~~~~~~~~~~~~~~~~~~~~~~~~~~~~~]{ \label{fig:X_mm-tr-2d}
\includegraphics[page=14]{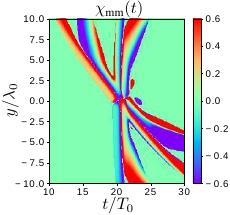}
}
\caption{ STV susceptibilities corresponding to the time-reversal oblique refraction transformation in Fig.~\ref{fig:metasurf_TR_ST}. The horizontal axis represents temporal evolution of the susceptibilities. (a) Electric susceptibility. (b) Magnetic susceptibility.}
\label{fig:susceptibilities-tr-2d}
\end{figure}

The corresponding STV susceptibilities, computed by \eqref{eq:X-mono-final}, are plotted in Fig.~\ref{fig:susceptibilities-tr-2d}, where the horizontal axis represents the time evolution of the susceptibilities. Note that the susceptibilities are at earlier times stronger at the upper half of the metasurface and at later times at its lower half. This gradual variation from the upper to the lower side of the metasurface, will be explained shortly in connection with the spatio-temporal evolution of the scattered fields. Note that, for transformations involving modulated multi-cycle pulses, the resulting STV susceptibility patterns will be quasi-periodic in both space and time. However, to produce susceptibilities with more easily recognizable features we opted for single cycle pulses in the examples presented in the paper.

\begin{figure}[ht!]
\centering
\subfloat[~~~~~~~~~~~~~~~~~~~~~~~~~~~~~~~~~~~~~~~~~]{ \label{fig:fdtd-reversal-field-before}
\includegraphics[page=15]{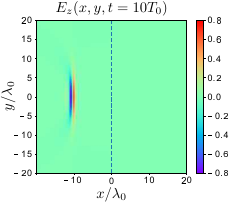}
}
\hfill
\subfloat[~~~~~~~~~~~~~~~~~~~~~~~~~~~~~~~~~~~~~~~~~]{ \label{fig:fdtd-reversal-field-after}
\includegraphics[page=16]{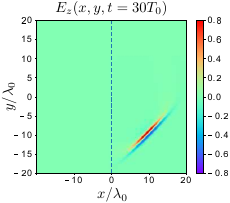}
}
\vfill
\subfloat[~~~~~~~~~~~~~~~~~~~~~~~~~~~~~~~~~~~~~~~~~]{ \label{fig:fdtd-reversal-field-spectrum-before}
\includegraphics[page=17]{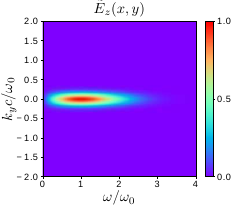}
}
\hfill
\subfloat[~~~~~~~~~~~~~~~~~~~~~~~~~~~~~~~~~~~~~~~~~]{ \label{fig:fdtd-reversal-field-spectrum-after}
\includegraphics[page=18]{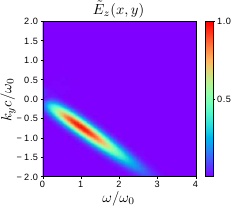}
}
\caption{Simulated FDTD fields for the time-reversal perfect refraction metasurface. (a),(b) Electric field corresponding to the incident and transmitted fields. (c),(d) Spatio-temporal spectrum of the incident and transmitted fields. The units in (a), (b) are V/m. The spectra in (c) and (d) are normalized.}
\label{fig:fdtd-tr}
\end{figure}

We simulated the scattering from this STV metasurface using the unstaggered FDTD scheme described in Sec.~\ref{Sec:unstaggered-fdtd}. Figures~\ref{fig:fdtd-reversal-field-before} and \ref{fig:fdtd-reversal-field-after} represent the fields incident on the metasurface and the scattered fields, respectively. The results perfectly match the specifications. These field profiles explain the peculiar space-time profile of the susceptibilities in Fig.~\ref{fig:susceptibilities-tr-2d}. The gradual variation of the susceptibilities from the upper to the lower side of the metasurface is due to the fact that, the upper part of the transmitted field is produced by the metasurface at earlier times, while the lower part of the transmitted field is produced at later times. Since at earlier and later times the incident pulse has a relatively small value on the metasurface, the susceptibilities have higher absolute values to compensate for the weak incident field. Note that at the center time, $t=21 T_0$, that corresponds to the arrival of the center of the incident pulse on the metasurface, the susceptibilities are significant all over the metasurface, as the metasurface has to attenuate the incident field at its upper and lower sections and transform the center region of the incident pulse around the center of the metasurface to the center region of the transmitted pulse. Animated FDTD results are provided as supplemental material~\cite{supplemental}.

The spatio-temporal spectrum of the incident and transmitted fields, obtained by space-time Fourier transformation, are presented in Figs.~\ref{fig:fdtd-reversal-field-spectrum-before} and \ref{fig:fdtd-reversal-field-spectrum-after}, respectively. The incident pulse is a wideband asymmetric pulse with spatial spectrum concentrated around $k_y=0$, corresponding to propagation along $x$. The transmitted spectrum corresponds to a pulse of identical bandwidth propagating at an oblique angle.

\subsection{Multifunction Time-Reversing Differentiating STV Metasurface}

\begin{figure}[ht!]
\centering
\subfloat[~~~~~~~~~~~~~~~~~~~~~~~~~~~~~~~~~~~~~~~~~]{
\includegraphics[page=19]{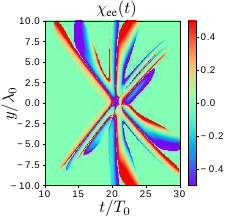}
}
\hfill
\subfloat[~~~~~~~~~~~~~~~~~~~~~~~~~~~~~~~~~~~~~~~~~]{
\includegraphics[page=20]{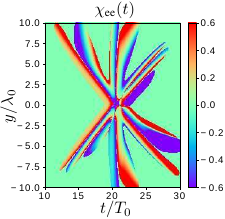}
}
\caption{STV susceptibilities corresponding to the multifunction time reversal differentiator metasurface in Fig.~\ref{fig:metasurf_TR_diff_ST}. The horizontal axis represents temporal evolution of the susceptibilities. (a) The electric susceptibility. (b) The magnetic susceptibility.}
\label{fig:susceptibilities-diff}
\end{figure}

Consider an asymmetric Gaussian pulse normally incident on the STV metasurface in Fig.~\ref{fig:metasurf_TR_diff_ST}. The metasurface simultaneously time-reverses and differentiates the incident pulse. It then refracts the time-reversed pulse at $-45$ degrees and the differentiated pulse at $+45$ degrees. The incident field is given in \eqref{eq:Ei-waveform-tr}. The specified transmission pulses can be obtained from \eqref{eq:Ei-waveform-tr} in a similar fashion, by applying first time-reversal or differentiation, then using  plane wave expansion or FDTD to obtain the fields in the $xy$ plane, and finally performing spatial rotations. The corresponding STV susceptibilities, computed using \eqref{eq:X-mono-final}, are plotted in Fig.~\ref{fig:susceptibilities-diff}.

\begin{figure}[ht!]
\centering
\subfloat[~~~~~~~~~~~~~~~~~~~~~~~~~~~~~~~~~~~~~~~~~]{ \label{fig:fdtd-diff-field-before}
\includegraphics[page=21]{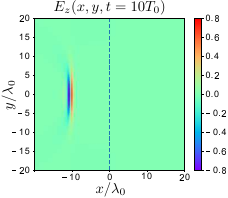}
}
\hfill
\subfloat[~~~~~~~~~~~~~~~~~~~~~~~~~~~~~~~~~~~~~~~~~]{ \label{fig:fdtd-diff-field-after}
\includegraphics[page=22]{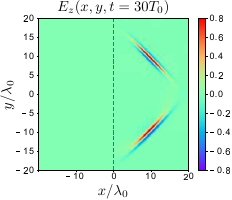}
}
\vfill
\subfloat[~~~~~~~~~~~~~~~~~~~~~~~~~~~~~~~~~~~~~~~~~]{ \label{fig:fdtd-diff-field-spectrum-before}
\includegraphics[page=23]{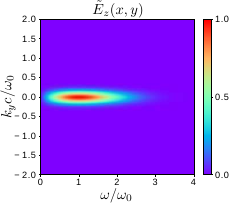}
}
\hfill
\subfloat[~~~~~~~~~~~~~~~~~~~~~~~~~~~~~~~~~~~~~~~~~]{ \label{fig:fdtd-diff-field-spectrum-after}
\includegraphics[page=24]{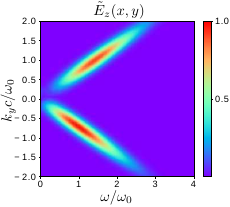}
}
\caption{FDTD results for the multifunction time reversing differentiating metasurface. (a),(b) Electric field corresponding to the incident and transmitted fields. (c),(d) Spatio-temporal spectrum of the incident and transmitted fields. The units in (a), (b) are V/m. The spectra in (c) and (d) are normalized to their maxima.}
\label{fig:fdtd-diff}
\end{figure}

The scattering response of the metasurface is modeled using the unstaggered FDTD method presented in Sec.~\ref{Sec:unstaggered-fdtd}. Figures~\ref{fig:fdtd-diff-field-before} and \ref{fig:fdtd-diff-field-after} represent the fields incident on the metasurface and the scattered fields, respectively. The results perfectly match the specifications. Animated FDTD results are provided a supplemental material~\cite{supplemental}. 

These field profiles explain the peculiar space-time profile of the susceptibilities in Fig.~\ref{fig:susceptibilities-diff}. The stronger susceptibilities at the upper part of the metasurface at early times and at the lower side of the metasurface at late times corresponds to the time reversal transformation. Similarly, the lower part of the differentiated pulse is produced at earlier times and its upper part is produced at later times, at the lower and upper parts of the metasurface, respectively, corresponding to the new branches in Figs.~\ref{fig:susceptibilities-diff} compared to the previous example. Finally, similar to the previous example, at the center time around $t=21 T_0$, that corresponds to the arrival of the center of the incident pulse on the metasurface, the susceptibilities are significant all over the metasurface, as the metasurface has to attenuate the incident field at its upper and lower sections and transform the center region of the incident pulse around the center of the metasurface to the center region of the transmitted time-reversed and differentiated pulses.

The spatio-temporal spectrum of the incident and transmitted fields are presented in Figs.~\ref{fig:fdtd-diff-field-spectrum-before} and \ref{fig:fdtd-diff-field-spectrum-after}, respectively. The incident pulse is a wideband asymmetric pulse with spatial spectrum concentrated around $k_y=0$, corresponding to propagation along $x$. The transmitted spectrum corresponds to two wideband pulses propagating at different oblique angles. Note that the differentiated pulse has a weaker frequency content at lower frequencies compared to the incident and time-reversed pulses, as differentiation is equivalent to multiplication by $j\omega$ in the frequency domain.

\section{Operation Limitation} \label{sec:discussion}
The synthesis technique presented in Sec.~\ref{sec:design} provides the required STV metasurface parameters for any given transformation. As examples, we presented metasurfaces that transform a pulse into its time-reversed or differentiated pulses. However, it should be noted that the synthesized metasurface only time reverses or differentiates the specified incidence pulse. For incident pulses slightly different from the specified one, the transformation would deteriorate, and for completely different incident pulses the behaviour of the metasurface would be erratic. This is an inherent limitation of GSTC designs and applies to both static and STV metasurfaces synthesized with this technique.

Another limitation of time varying structures is that their transformations are \emph{(time-)shift variant}, i.e., the transformation is different for input pulses shifted in time. As a consequence, the incident pulses in all the examples presented in Sec.~\ref{sec:examples} must be synchronized with the metasurface. 

Thus, it would be more appropriate to regard the presented transformations in Sec.~\ref{sec:examples} as pulse-shaping transformations using coherent STV metasurfaces, rather than time-reversal or differentiation in the broad sense. However, it should be noted that pulse coherency is now a well established technique that is present in most digital radio and optical communication systems, and can be achieved using phase-locked loops or similar feedback systems. The presented STV metasurfaces may find applications in radio or optical detection systems where for example binary state incident pulses are transformed to pulse shapes that are more easily detectable by the available optical or radio detectors. For example the metasurface could transform information encoded in complex pulse shapes into on-off keying logic.

Finally, it is worth mentioning a paradoxical transformation where a \emph{zero} incident field is specified to produce \emph{non-zero} reflected and transmitted fields. As with previous examples, one may proceed to find corresponding STV susceptibilities using \eqref{eq:X-mono-final}. It may appear then, that the synthesized metasurface must produce the specified scattered fields out of \emph{nothing}. However, this is not the case. It may be confirmed using \eqref{eq:time-only-1d-gstc-mono-iso-PM} as well as using FDTD analysis, that for zero incident field the scattered fields are always zero. In this exceptional case, the specified scattered fields constitute a \emph{mode} of the metasurface that could potentially be excited with a proper incident field. A similar example is a guided or surface-wave transformation in a time invariant metasurface. It is common practice to define metasurface susceptibilities that support surface waves in the absence of any incident fields. However, such surface waves are merely \emph{modes} of the metasurface and will not be generated unless the metasurface is excited with proper incident fields.

\section{Conclusions} \label{sec:conclusions}
STV metasurfaces have been proposed for simultaneously controling the temporal and spatial spectra of electromagnetic waves. A systematic technique based on time-varying GSTCs has been presented for the design of such metasurfaces. Analytical and numerical techniques for analysis of such metasurfaces, has been presented, and used to demonstrate operations such as time-reversal, differentiation and perfect pulse refraction. The proposed metasurfaces may find applications in pulse shaping devices. Moreover, they may extend the functionality of static metasurfaces by lifting physical limits imposed on LTI systems. STV metasurfaces naturally break Lorentz reciprocity and therefore may find applications in magnetless non-reciprocal metasurfaces as well.

\appendices

\section{Analytic solution of 1D time-varying metasurface scattering problems}  \label{appendix:analytic-1d}
This appendix derives the analytical solution to the first-order differential equations~\eqref{eq:time-only-1d-gstc-mono-iso-PM}. These equations can be expressed in the general form

\begin{equation} \label{eq:appendix-analytic-ode-with-singular-point}
\frac{d}{d t}f\left(t\right)+a\left(t\right)f\left(t\right)/\left(t-t_{0}\right)=g\left(t\right),
\end{equation}
where $f$ represents $P_\text{t}$ or $M_\text{t}$ and $t_0$ is a zero of $\chi_\text{ee}(t)$ or $\chi_\text{mm}(t)$, and $g(t)$ is proportional to $E^\text{i}(t)$. 

Equation~\eqref{eq:appendix-analytic-ode-with-singular-point} can be solved by transforming its left hand side into a perfect derivative~\cite{coddington1955theory}. This may be accomplished by first multiplying both sides by the still unknown function $\theta(t)$, i.e.,
\begin{equation} \label{eq:appendix-analytic-ode-with-singular-point-multiply-theta}
\theta(t)\frac{d}{d t}f\left(t\right)+\theta(t)a\left(t\right)f\left(t\right)/\left(t-t_{0}\right)=\theta(t)g\left(t\right),
\end{equation}
and next noting that the left hand sides makes a perfect derivative $\frac{d}{d t}\left[\theta(t)f\left(t\right)\right]$, transforming \eqref{eq:appendix-analytic-ode-with-singular-point-multiply-theta} into 

\begin{equation} \label{eq:appendix-analytic-ode-with-singular-point-perfect-derivative-lhs}
\frac{d}{d t}\left[\theta(t)f\left(t\right)\right] = \theta(t)g\left(t\right),
\end{equation}
if
\begin{equation} \label{eq:appendix-analytic-theta-integral}
\theta\left(t\right)=\exp\left(\int a\left(t\right)/\left(t-t_{0}\right)dt\right).
\end{equation}
Integrating \eqref{eq:appendix-analytic-ode-with-singular-point-perfect-derivative-lhs} yields then

\begin{equation} \label{eq:appendix-analytic-f-solution}
f\left(t\right)=\left[\int\theta\left(t\right)g\left(t\right)dt+c\right]/\theta\left(t\right),
\end{equation}
where $c$ is an arbitrary constant to be fixed by the initial condition.

Using integration by parts \eqref{eq:appendix-analytic-theta-integral} becomes
 
\begin{align} \label{eq:appendix-analytic-theta-expanded}
\theta\left(t\right) &= \exp\left[a\left(t\right)\ln|t-t_{0}|-\int\ln|t-t_{0}|a^{\prime}\left(t\right)dt\right] \\\nonumber
&= |t-t_{0}|^{a\left(t\right)}/\exp\left[\int\ln|t-t_{0}|a^{\prime}\left(t\right)dt\right].
\end{align}
Close to the singularity $t=t_0$ we may use L'H\^opital rule, which changes \eqref{eq:appendix-analytic-f-solution} to

\begin{equation} \label{eq:appendix-analytic-f-singularity}
f\left(t\right)=\theta\left(t\right)g\left(t\right)/\theta^{\prime}\left(t\right)=\left(t-t_{0}\right)g\left(t\right)/a\left(t\right),
\end{equation}
where, the last equality uses

\begin{equation} \label{eq:appendix-analytic-theta_prime_over_theta}
\theta^{\prime}\left(t\right)/\theta\left(t\right)=a\left(t\right)/\left(t-t_{0}\right),
\end{equation}
which is obtained from \eqref{eq:appendix-analytic-theta-integral}.

In summary $f(t)$ is provided by \eqref{eq:appendix-analytic-f-solution} for $t \neq t_0$ and by  \eqref{eq:appendix-analytic-f-singularity} for $t = t_0$, which leads to analytical expressions in some simple cases.

Note that $\chi_\text{ee}(t)$ or $\chi_\text{mm}(t)$ may have multiple zeros in general. In that case the time axis can be subdivided into contiguous intervals each containing only one zero, and the process outlined in this section can be repeated for each interval.

\section{FDTD Discretization of Maxwell Equations on The Unstaggered Grid} \label{appendix:fdtd-unstaggered-maxwell}
This appendix presents derivation details for the unstaggered FDTD scheme, used in this paper, for both 1D and 2D uniform Cartesian grids.

\subsection{1D FDTD Equations}
\begin{figure}[ht!]
\centering
\includegraphics[page=25]{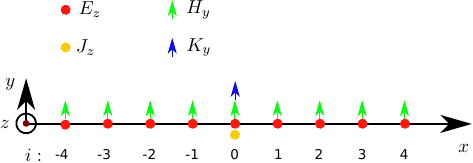}
\caption{Location of the electric and magnetic field components in the 1D unstaggered grid. The grid nodes are placed at $x_i = i \Delta x$ with $\Delta x = \text{const.}$, and where $i$ is an integer. The metasurface is represented by the equivalent electric and magnetic surface current densities at $x=0$.}
\label{fig:fdtd-unstaggered-1d-EH-locs}
\end{figure}

Figure~\ref{fig:fdtd-unstaggered-1d-EH-locs} shows location of the relevant field components in the 1D unstaggered grid. The electric and magnetic fields as well as all the other electromagnetic quantities such as the current densities, are located at the same nodes. In a uniform grid these nodes are equidistantly located at the grid points $x_i = i\Delta x$ as shown in Fig.~\ref{fig:fdtd-unstaggered-1d-EH-locs}. For simplicity we assume the electric and magnetic fields are along the $z$ and $y$ directions, respectively, and are propagating in free space. The metasurface is represented by equivalent electric and magnetic surface current densities at $x=0$. We represent the spatial and temporal indices by subscript and superscripts, respectively. For instance $E^n_i = E(i\Delta x, n\Delta t)$, assuming the spatial and temporal resolutions $\Delta x$ and $\Delta t$, respectively.

The corresponding 1D Maxwell equations read

\begin{subequations}
\begin{align}
-\frac{\partial}{\partial x}E_{z}\left(x,t\right) &= -\mu_{0}\frac{\partial}{\partial t}H_{y}\left(x,t\right)-K_{y}\left(x,t\right), \\
\frac{\partial}{\partial x}H_{y}\left(x,t\right) &= \epsilon_{0}\frac{\partial}{\partial t}E_{z}\left(x,t\right)+J_{z}\left(x,t\right),
\end{align} \label{eq:maxwell-1d}
\end{subequations}
where $J_z$ and $K_y$ are the electric and magnetic current densities, respectively. Applying the central difference schemes in \eqref{eq:fdtd-dxdt-central-difference} discretizes \eqref{eq:maxwell-1d} as

\begin{subequations}
\begin{align}
\frac{E_{i+1}^{n}-E_{i-1}^{n}}{2\Delta x} &= \mu_0\frac{H_{i}^{n+1}-H_{i}^{n-1}}{2\Delta t}+K_{i}^{n}, \\
\frac{H_{i+1}^{n}-H_{i-1}^{n}}{2\Delta x} &= \epsilon_0\frac{E_{i}^{n+1}-E_{i}^{n-1}}{2\Delta t}+J_{i}^{n}.
\end{align} \label{eq:maxwell-1d-fd}
\end{subequations}
The electric and magnetic fields can be explicitly updated at the time step $n+1$, using the fields at previous time steps, as follows

\begin{subequations}
\begin{align}
H_{i}^{n+1} &= H_{i}^{n-1}+\frac{\Delta t}{\mu_{0}\Delta x}\left(E_{i+1}^{n}-E_{i-1}^{n}\right)-\frac{2\Delta t}{\mu_{0}}K_{i}^{n}, \\
E_{i}^{n+1} &= E_{i}^{n-1}+\frac{\Delta t}{\epsilon_{0}\Delta x}\left(H_{i+1}^{n}-H_{i-1}^{n}\right)-\frac{2\Delta t}{\epsilon_{0}}J_{i}^{n},
\end{align} \label{eq:maxwell-1d-fdtd-update-equations}
\end{subequations}

\noindent
where $K_{i}^{n}$ and $J_{i}^{n}$ are zero everywhere except on the metasurface.

The metasurface polarization densities are updated using~\eqref{eq:constitut-mono-iso}, starting with

\begin{subequations}
\begin{align}
P_0^n &= \epsilon_0 \chi_\text{ee}^n E_\text{av}^n = \epsilon_0 \chi_\text{ee}^n E_0^n, \\
M_0^n &= \chi_\text{mm}^n H_\text{av}^n = \chi_\text{mm}^n H_0^n,
\end{align} \label{eq:maxwell-1d-PM-0}
\end{subequations}
where $\chi_\text{ee}^n = \chi_\text{ee} (n \Delta t)$, $\chi_\text{mm}^n = \chi_\text{mm} (n \Delta t)$. Finally, the equivalent current densities can be computed using
\begin{subequations}
\begin{align}
\mathbf{J}_\text{eq} &= \frac{\partial}{\partial t} \mathbf{P}, \\
\mathbf{K}_\text{eq} &= \mu_0\frac{\partial}{\partial t} \mathbf{M},
\end{align} \label{eq:maxwell-1d-PM-0}
\end{subequations}
\noindent
which can be updated using first-order or second-order backward-difference schemes. For the former case the resulting equations take the form

\begin{subequations}
\begin{align}
J_0^n &= \frac{P_0^n - P_0^{n-1}}{\Delta t}, \\
K_0^n &= \mu_0\frac{M_0^n - M_0^{n-1}}{\Delta t},
\end{align} \label{eq:maxwell-1d-PM-0}
\end{subequations}
which can be written explicitly in terms of the susceptibilities and the electromagnetic fields as

\begin{subequations}
\begin{align}
J_0^n &= \epsilon_0 \frac{\chi_\text{ee}^n E_0^n - \chi_\text{ee}^{n-1} E_0^{n-1}}{\Delta t}, \\
K_0^n &= \mu_0 \frac{\chi_\text{mm}^n H_0^n - \chi_\text{mm}^{n-1} H_0^{n-1}}{\Delta t}.
\end{align} \label{eq:maxwell-1d-PM-0}
\end{subequations}

\subsection{2D FDTD Equations}
\begin{figure}[ht!]
\centering
\includegraphics[page=26]{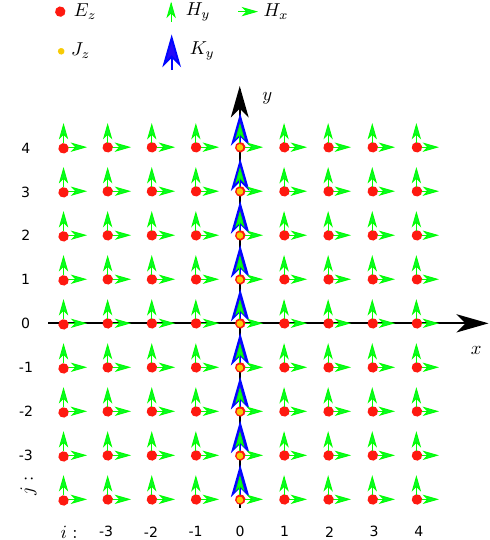}
\caption{Location of the electric and magnetic field components in the 2D unstaggered grid. The grid nodes are placed at $(x_i, y_j) = (i \Delta x, j\Delta y)$ with $\Delta x, \Delta y = \text{const.}$, and where $i$ and $j$ are integers. The metasurface is placed in the plane $x=0$ and is represented by the equivalent electric and magnetic surface current densities.}
\label{fig:fdtd-unstaggered-2d-EH-locs}
\end{figure}

Figure~\ref{fig:fdtd-unstaggered-2d-EH-locs} shows the location of the different field components on a 2D unstaggered grid. The electric and magnetic fields as well as all the other electromagnetic quantities such as the current densities, are located at the same nodes. In a uniform grid, these nodes are equidistantly located at the grid points $(x_i, y_j) = (i\Delta x, j\Delta y)$ ($\Delta x, \Delta y = \text{const.}$) as shown in Fig.~\ref{fig:fdtd-unstaggered-1d-EH-locs}. For brevity we consider only the transverse electric (TE) mode, where the electric field is along the $z$ direction and the magnetic field is in the $xy$-plane, and are propagating in free space. The results can be straightforwardly extended to more general cases. The metasurface is represented by equivalent electric and magnetic surface current densities in the plane $x=0$. We represent spatial and temporal indices by subscript and superscripts, respectively. For instance $E^n_{ij} = E(i\Delta x, j\Delta y, n\Delta t)$, assuming a grid with spatial and temporal resolutions $(\Delta x, \Delta y)$ and $\Delta t$, respectively.

The corresponding 2D Maxwell equations read

\begin{subequations}
\begin{align}
&-\frac{\partial}{\partial x}E_{z}\left(x,y,t\right) = -\mu_{0}\frac{\partial}{\partial t}H_{y}\left(x,y,t\right)-K_{y}\left(x,y,t\right), \\
&\frac{\partial}{\partial y}E_{z}\left(x,y,t\right)=-\mu_{0}\frac{\partial}{\partial t}H_{x}\left(x,y,t\right), \\
&\frac{\partial}{\partial x}H_{y}\left(x,y,t\right)-\frac{\partial}{\partial y}H_{x}\left(x,y,t\right) = \epsilon_{0}\frac{\partial}{\partial t}E_{z}\left(x,y,t\right) \\\nonumber
&+J_{z}\left(x,y,t\right)
\end{align} \label{eq:maxwell-2d}
\end{subequations}
where $J_z \hat{\mathbf{z}}$ and  $K_x \hat{\mathbf{x}} + K_y \hat{\mathbf{y}}$ are the equivalent electric and magnetic current densities, respectively. Applying the central difference schemes in \eqref{eq:fdtd-dxdt-central-difference} discretizes \eqref{eq:maxwell-2d} as

\begin{subequations}
\begin{align}
&\frac{{_{z}}E_{i+1,j}^{n}-{_{z}}E_{i-1,j}^{n}}{2\Delta x}=\mu_{0}\frac{_{y}H_{ij}^{n+1}-{_{y}}H_{ij}^{n-1}}{2\Delta t}+K_{ij}^{n} \\
&\frac{{_{z}}E_{i,j+1}^{n}-{_{z}}E_{i,j-1}^{n}}{2\Delta y}=-\mu_{0}\frac{_{x}H_{ij}^{n+1}-{_{x}}H_{ij}^{n-1}}{2\Delta t} \\
&\frac{_{y}H_{i+1,j}^{n}-{_{y}}H_{i-1,j}^{n}}{2\Delta x}-\frac{_{x}H_{i,j+1}^{n}-{_{x}}H_{i,j-1}^{n}}{2\Delta y} = \\\nonumber
&\epsilon_{0}\frac{{_{z}}E_{ij}^{n+1}-{_{z}}E_{ij}^{n-1}}{2\Delta t}+J_{ij}^{n}
\end{align} \label{eq:maxwell-2d-fd}
\end{subequations}
where the $x, y, z$ vector component letters are placed before the fields, as in $_z E_{ij}^n$, in order to avoid mixing them with the grid indices $i,j,n$. The electric and magnetic fields can be explicitly updated at the time step $n + 1$, using the fields at previous time steps, as follows

\begin{subequations}
\begin{align}
&_{z}E_{ij}^{n+1}={}_{z}E_{ij}^{n-1}+\frac{\Delta t}{\epsilon_{0}\Delta x}\left(_{y}H_{i+1,j}^{n}-{}_{y}H_{i-1,j}^{n}\right)\\\nonumber
&-\frac{\Delta t}{\epsilon_{0}\Delta y}\left(_{x}H_{i,j+1}^{n}-{}_{x}H_{i,j-1}^{n}\right)-\frac{2\Delta t}{\epsilon_{0}}J_{ij}^{n} \\
&_{x}H_{ij}^{n+1}={}_{x}H_{ij}^{n-1}-\frac{\Delta t}{\mu_{0}\Delta y}\left(_{z}E_{i,j+1}^{n}-{}_{z}E_{i,j-1}^{n}\right) \\
&_{y}H_{ij}^{n+1}={}_{y}H_{ij}^{n-1}+\frac{\Delta t}{\mu_{0}\Delta x}\left(_{z}E_{i+1,j}^{n}-{}_{z}E_{i-1,j}^{n}\right)-\frac{2\Delta t}{\mu_{0}}K_{ij}^{n}
\end{align} \label{eq:maxwell-2d-fdtd}
\end{subequations}

The metasurface polarization densities are updated using~\eqref{eq:constitut-mono-iso}, starting with

\begin{subequations}
\begin{align}
P_{0j}^n &= \epsilon_0 \left(\chi_\text{ee}\right)_j^n {}_z E_\text{av}^n = \epsilon_0 \left(\chi_\text{ee}\right)_j^n {}_z E_{0j}^n, \\
M_{0j}^n &= \left(\chi_\text{mm}\right)_j^n {}_y H_\text{av}^n = \left(\chi_\text{mm}\right)_j^n {}_y H_{0j}^n,
\end{align} \label{eq:maxwell-2d-PM-0}
\end{subequations}
where $\left(\chi_\text{ee}\right)_j^n = \chi_\text{ee} (j \Delta y, n \Delta t)$, $\left(\chi_\text{mm}\right)_j^n = \chi_\text{mm} (j \Delta y, n \Delta t)$. Finally, the equivalent current densities can be computed using \eqref{eq:maxwell-1d-PM-0}, which can be updated using first-order or second-order backward-difference schemes. For the former case the resulting equations take the form

\begin{subequations}
\begin{align}
J_{0j}^n &= \frac{P_{0j}^n - P_{0j}^{n-1}}{\Delta t}, \\
K_{0j}^n &= \mu_0\frac{M_{0j}^n - M_{0j}^{n-1}}{\Delta t},
\end{align} \label{eq:maxwell-2d-PM-0}
\end{subequations}
which can be written explicitly in terms of the susceptibilities and the electromagnetic fields as

\begin{subequations}
\begin{align}
J_{0j}^n &= \epsilon_0 \frac{\left(\chi_\text{ee}\right)_j^n {}_z E_{0j}^n - \left(\chi_\text{ee}\right)_j^{n-1} {}_z E_{0j}^{n-1}}{\Delta t}, \\
K_{0j}^n &= \mu_0 \frac{\left(\chi_\text{mm}\right)_j^n {}_y H_{0j}^n - \left(\chi_\text{mm}\right)_j^{n-1} {}_y H_{0j}^{n-1}}{\Delta t}.
\end{align} \label{eq:maxwell-2d-PM-0}
\end{subequations}

\bibliographystyle{IEEEtran}
\bibliography{ReferenceList2_abbr}

\begin{thebibliography}{10}
\providecommand{\url}[1]{#1}
\csname url@samestyle\endcsname
\providecommand{\newblock}{\relax}
\providecommand{\bibinfo}[2]{#2}
\providecommand{\BIBentrySTDinterwordspacing}{\spaceskip=0pt\relax}
\providecommand{\BIBentryALTinterwordstretchfactor}{4}
\providecommand{\BIBentryALTinterwordspacing}{\spaceskip=\fontdimen2\font plus
\BIBentryALTinterwordstretchfactor\fontdimen3\font minus
  \fontdimen4\font\relax}
\providecommand{\BIBforeignlanguage}[2]{{%
\expandafter\ifx\csname l@#1\endcsname\relax
\typeout{** WARNING: IEEEtran.bst: No hyphenation pattern has been}%
\typeout{** loaded for the language `#1'. Using the pattern for}%
\typeout{** the default language instead.}%
\else
\language=\csname l@#1\endcsname
\fi
#2}}
\providecommand{\BIBdecl}{\relax}
\BIBdecl

\bibitem{holloway2012overview}
C.~L. Holloway, E.~F. Kuester, J.~A. Gordon, J.~O'Hara, J.~Booth, and D.~R.
  Smith, ``An overview of the theory and applications of metasurfaces: The
  two-dimensional equivalents of metamaterials,'' \emph{IEEE Antennas Propag.
  Mag.}, vol.~54, no.~2, pp. 10--35, 2012.

\bibitem{achouri2015general}
K.~Achouri, M.~A. Salem, and C.~Caloz, ``General metasurface synthesis based on
  susceptibility tensors,'' \emph{IEEE Trans. Antennas Propag.}, vol.~63,
  no.~7, pp. 2977--2991, 2015.

\bibitem{yu2014flat}
N.~Yu and F.~Capasso, ``Flat optics with designer metasurfaces,'' \emph{Nat.
  Mater.}, vol.~13, no.~2, p. 139, 2014.

\bibitem{zheng2015metasurface}
G.~Zheng, H.~M{\"u}hlenbernd, M.~Kenney, G.~Li, T.~Zentgraf, and S.~Zhang,
  ``Metasurface holograms reaching 80\% efficiency,'' \emph{Nat. Nanotechnol.},
  vol.~10, no.~4, p. 308, 2015.

\bibitem{karimi2014generating}
E.~Karimi, S.~A. Schulz, I.~De~Leon, H.~Qassim, J.~Upham, and R.~W. Boyd,
  ``Generating optical orbital angular momentum at visible wavelengths using a
  plasmonic metasurface,'' \emph{Light. Sci. \& Appl.}, vol.~3, no.~5, p. e167,
  2014.

\bibitem{pfeiffer2014high}
C.~Pfeiffer, C.~Zhang, V.~Ray, L.~J. Guo, and A.~Grbic, ``High performance
  bianisotropic metasurfaces: asymmetric transmission of light,'' \emph{Phys.
  Rev. Lett.}, vol. 113, no.~2, p. 023902, 2014.

\bibitem{achouri2016metasurfaceremote}
K.~Achouri, G.~Lavigne, M.~A. Salem, and C.~Caloz, ``Metasurface spatial
  processor for electromagnetic remote control,'' \emph{IEEE Trans. Antennas
  Propag.}, vol.~64, no.~5, pp. 1759--1767, 2016.

\bibitem{chen2017simultaneous}
L.~Chen, K.~Achouri, E.~Kallos, and C.~Caloz, ``Simultaneous enhancement of
  light extraction and spontaneous emission using a partially reflecting
  metasurface cavity,'' \emph{Phys. Rev. A}, vol.~95, no.~5, p. 053808, 2017.

\bibitem{achouri2017solarsail}
K.~Achouri and C.~Caloz, ``Metasurface solar sail for flexible radiation
  pressure control,'' \emph{arXiv Prepr. arXiv:1710.02837}, 2017.

\bibitem{bode1945network}
H.~W. Bode \emph{et~al.}, ``Network analysis and feedback amplifier design,''
  1945.

\bibitem{shlivinski2018paradigm}
A.~Shlivinski and Y.~Hadad, ``A paradigm for instantaneously-wideband impedance
  matching by temporal switching of transmission line parameters,'' \emph{arXiv
  Prepr. arXiv:1805.03704}, 2018.

\bibitem{shaltout2015time}
A.~Shaltout, A.~Kildishev, and V.~Shalaev, ``Time-varying metasurfaces and
  lorentz non-reciprocity,'' \emph{Opt. Mater. Expr.}, vol.~5, no.~11, pp.
  2459--2467, 2015.

\bibitem{hadad2015space}
Y.~Hadad, D.~Sounas, and A.~Al\`u, ``Space-time gradient metasurfaces,''
  \emph{Phys. Rev. B}, vol.~92, no.~10, p. 100304, 2015.

\bibitem{caloz2018nonreciprocity}
C.~Caloz, A.~Al\`u, S.~Tretyakov, D.~Sounas, K.~Achouri, and Z.-L. Deck-Leger,
  ``Electromagnetic nonreciprocity,'' \emph{Phys. Rev. Appl.}, 2018, to be
  published.

\bibitem{liu2018huygens}
M.~Liu, D.~A. Powell, Y.~Zarate, and I.~V. Shadrivov, ``Huygens’ metadevices
  for parametric waves,'' \emph{Phys. Rev. X}, vol.~8, no.~3, p. 031077, 2018.

\bibitem{yu2009complete}
Z.~Yu and S.~Fan, ``Complete optical isolation created by indirect interband
  photonic transitions,'' \emph{Nat. Photon.}, vol.~3, no.~2, p.~91, 2009.

\bibitem{sounas2013giant}
D.~L. Sounas, C.~Caloz, and A.~Al\`u, ``Giant non-reciprocity at the
  subwavelength scale using angular momentum-biased metamaterials,'' \emph{Nat.
  Commun.}, vol.~4, p. 2407, 2013.

\bibitem{chamanara2017optical}
N.~Chamanara, S.~Taravati, Z.-L. Deck-L{\'e}ger, and C.~Caloz, ``Optical
  isolation based on space-time engineered asymmetric photonic band gaps,''
  \emph{Phys. Rev. B}, vol.~96, no.~15, p. 155409, 2017.

\bibitem{taravati2017nonreciprocal}
S.~Taravati, N.~Chamanara, and C.~Caloz, ``Nonreciprocal electromagnetic
  scattering from a periodically space-time modulated slab and application to a
  quasisonic isolator,'' \emph{Physical Review B}, vol.~96, no.~16, p. 165144,
  2017.

\bibitem{neto2009electronic}
A.~C. Neto, F.~Guinea, N.~M. Peres, K.~S. Novoselov, and A.~K. Geim, ``The
  electronic properties of graphene,'' \emph{Rev. Mod. Phys.}, vol.~81, no.~1,
  p. 109, 2009.

\bibitem{geim2010rise}
A.~K. Geim and K.~S. Novoselov, ``The rise of graphene,'' in \emph{Nanoscience
  and Technology: A Collection of Reviews from Nature Journals}.\hskip 1em plus
  0.5em minus 0.4em\relax World Scientific, 2010, pp. 11--19.

\bibitem{chamanara2013coupler}
N.~Chamanara, D.~Sounas, and C.~Caloz, ``Non-reciprocal magnetoplasmon graphene
  coupler,'' \emph{Opt. Expr.}, vol.~21, no.~9, pp. 11\,248--11\,256, 2013.

\bibitem{chamanara2012transparentgraphene}
N.~Chamanara, D.~Sounas, T.~Szkopek, and C.~Caloz, ``Optically transparent and
  flexible graphene reciprocal and nonreciprocal microwave planar components,''
  \emph{IEEE Microw. Wirel. Compon. Lett.}, vol.~22, no.~7, pp. 360--362, 2012.

\bibitem{chamanara2016grapheneTEplasmon}
N.~Chamanara and C.~Caloz, ``Graphene transverse electric surface plasmon
  detection using nonreciprocity modal discrimination,'' \emph{Phys. Rev. B},
  vol.~94, no.~7, p. 075413, 2016.

\bibitem{chamanara2015fundamentals}
------, ``Fundamentals of graphene magnetoplasmons: principles, structures and
  devices,'' \emph{FERMAT}, vol.~1, no.~3, p.~3, 2015.

\bibitem{kinsey2015epsilon}
N.~Kinsey, C.~DeVault, J.~Kim, M.~Ferrera, V.~Shalaev, and A.~Boltasseva,
  ``Epsilon-near-zero al-doped zno for ultrafast switching at telecom
  wavelengths,'' \emph{Opt.}, vol.~2, no.~7, pp. 616--622, 2015.

\bibitem{ferrera2017dynamic}
M.~Ferrera, N.~Kinsey, A.~Shaltout, C.~DeVault, V.~Shalaev, and A.~Boltasseva,
  ``Dynamic nanophotonics,'' \emph{JOSA B}, vol.~34, no.~1, pp. 95--103, 2017.

\bibitem{achouri2018design}
K.~Achouri and C.~Caloz, ``Design, concepts, and applications of
  electromagnetic metasurfaces,'' \emph{Nanophotonics}, vol.~7, no.~6, pp.
  1095--1116, 2018.

\bibitem{chamanara2017efficient}
N.~Chamanara, K.~Achouri, and C.~Caloz, ``Efficient analysis of metasurfaces in
  terms of spectral-domain gstc integral equations,'' \emph{IEEE Trans.
  Antennas Propag.}, vol.~65, no.~10, pp. 5340--5347, 2017.

\bibitem{chamanara2016exact}
N.~Chamanara, Y.~Vahabzadeh, K.~Achouri, and C.~Caloz, ``Exact polychromatic
  metasurface design: The gstc approach,'' in \emph{Adv. Electromagn. Mater.
  Microw. Opt. (METAMATERIALS), 2016 10th Int. Congr.}\hskip 1em plus 0.5em
  minus 0.4em\relax IEEE, 2016, pp. 91--93.

\bibitem{schwinger1998classical}
J.~Schwinger, L.~L. DeRaad~Jr, K.~Milton, and W.-y. Tsai, \emph{Classical
  electrodynamics}.\hskip 1em plus 0.5em minus 0.4em\relax Westview Press,
  1998.

\bibitem{jackson2012classical}
J.~D. Jackson, \emph{Classical electrodynamics}.\hskip 1em plus 0.5em minus
  0.4em\relax John Wiley \& Sons, 2012.

\bibitem{ishimaru2017electromagnetic}
A.~Ishimaru, \emph{Electromagnetic wave propagation, radiation, and scattering:
  from fundamentals to applications}.\hskip 1em plus 0.5em minus 0.4em\relax
  John Wiley \& Sons, 2017.

\bibitem{idemen1987boundary}
M.~Idemen and A.~H. Serbest, ``Boundary conditions of the electromagnetic
  field,'' \emph{Electron. Lett.}, vol.~23, no.~13, pp. 704--705, 1987.

\bibitem{taflove2005computational}
A.~Taflove and S.~C. Hagness, \emph{Computational electrodynamics: the
  finite-difference time-domain method}.\hskip 1em plus 0.5em minus 0.4em\relax
  Artech house, 2005.

\bibitem{vahabzadeh2018generalized}
Y.~Vahabzadeh, N.~Chamanara, and C.~Caloz, ``Generalized sheet transition
  condition fdtd simulation of metasurface,'' \emph{IEEE Trans. Antennas
  Propag.}, vol.~66, no.~1, pp. 271--280, 2018.

\bibitem{vahabzadeh2018computational}
Y.~Vahabzadeh, N.~Chamanara, K.~Achouri, and C.~Caloz, ``Computational analysis
  of metasurfaces,'' \emph{IEEE J. Multiscale Multiphysics Comput. Tech.},
  vol.~3, pp. 37--49, 2018.

\bibitem{gilles2000comparison}
L.~Gilles, S.~Hagness, and L.~V{\'a}zquez, ``Comparison between staggered and
  unstaggered finite-difference time-domain grids for few-cycle temporal
  optical soliton propagation,'' \emph{J. Comput. Phys.}, vol. 161, no.~2, pp.
  379--400, 2000.

\bibitem{janaswamy1997unstaggered}
R.~Janaswamy and Y.~Liu, ``An unstaggered colocated finite-difference scheme
  for solving time-domain maxwell's equations in curvilinear coordinates,''
  \emph{IEEE Trans. Antennas Propag.}, vol.~45, no.~11, pp. 1584--1591, 1997.

\bibitem{stewart2018finite}
S.~A. Stewart, T.~J. Smy, and S.~Gupta, ``Finite-difference time-domain
  modeling of space--time-modulated metasurfaces,'' \emph{IEEE Trans. Antennas
  Propag.}, vol.~66, no.~1, pp. 281--292, 2018.

\bibitem{liu1996fourier}
Y.~Liu, ``Fourier analysis of numerical algorithms for the maxwell equations,''
  \emph{J. Comput. Phys.}, vol. 124, no.~2, pp. 396--416, 1996.

\bibitem{supplemental}
\href{https://drive.google.com/file/d/1WdiX62Wr32QujbAPxpoR5frKCVMRphkQ/view?usp=sharing}{Supplemental
  material}.

\bibitem{coddington1955theory}
E.~A. Coddington and N.~Levinson, \emph{Theory of ordinary differential
  equations}.\hskip 1em plus 0.5em minus 0.4em\relax Tata McGraw-Hill
  Education, 1955.

\end{thebibliography}


%\begin{IEEEbiography}{Nima Chamanara}
%\end{IEEEbiography}
%\begin{IEEEbiographynophoto}{Karim Achouri}
%% insert where needed to balance the two columns on the last page with
%%\newpage


%\vfill


\end{document}